# Effect of backing thickness on determination of the phase in neutron reflectometry by variation of backing

S.F. Masoudi[a], A. Pazirandeh[a], and G.R. Jafari[b]

[a]Department of Physics, Tehran University, P.O. Box 1943-19395, Tehran, Iran
[b]Department of physics, Sharif University of Technology, P.O. Box 11365-9161, Tehran, Iran

The determination of density profiles with knowing the phase information of complex reflection coefficient for neutron specularly reflected from a film, yields unique results. Recently it has been shown that the phase can be determined by using controlled variation of the scattering length density of the fronting (incident) and/or backing (substrate) medium instead of reference layers of finite thickness. This method is applicable under the simplifying assumption that the backing is infinitely thick (semi-infinite substrate). By this assumption the reflected beams from the end side of the backing is neglected, which appears reasonable since in most cases absorption will damp out the neutron current before it has reached the end side. But for weakly absorbing backing, the reflection from the end side may be considerable. So backing must be considered as a thick matter. We show that for this kind of backing, using method of variation of the backing, as an example of variation of the surroundings, leads to completely wrong answer in determination of the phase.

## I. INTRODUCTION

Neutron specular reflectometry is a potentially powerful method to probe many surface and interfacial structures, in fields as diverse as polymers and magnetism.[1,2] The measurement of the reflectivity, R(q), from a sample as a function of the perpendicular component of the incident wave vector q=2π sin θ/λ, with λ the neutron wavelength and θ the reflection angle, provides information to extract the scattering length density (SLD) depth profile of the sample. In reflection experiment only the square of the complex reflection coefficient r(q), is measured so like any other scattering technique the phase of reflection is lost. In the absence of the phase, generally least-squares methods[3] are used to extract the SLD profile, but in general more than one SLD may be found to correspond to the same reflectivity.[4] Given the phase, it is possible to solve the one-dimensional inverse-scattering problem directly to obtain a unique SLD depth profile.[5] If the SLD profiles are nowhere negative, the analytic properties of r(q) ensure that the inversion is unique and that either the real or imaginary part of r(q) is sufficient data.[6]

Several methods for measuring phase have been developed for neutron specular reflection.[7-12] Among these methods, the reference layer method, which introduced in Ref. 9, Seems the best one because of its application in experiment. Instead of the variation of buried layers of finite thickness, the variation of the uniform fronting (incident) and/or backing (transmitting media) can be used in some systems in which the SLD profile of the film under study be stable against variation of the SLD of the surround. In this method it is common to consider the backing as a semi infinite matter like in "vacuum fronting, variable backing" method.[13] By this simplifying assumption, only a single scalar value of the SLD is needed for obtaining the phase, but the reflection from the back side of the backing is neglected. This assumption is good and reasonable in most cases in which the absorption is not weak. For weakly absorbing materials like, e.g., silicon, ignoring the waves reflected from the end side of the backing, that have appreciable effects on the reflectivity, leads to a wrong interpretation of measured reflectivities.[14]

The aim of the present paper is to investigate these effects on "variation of surrounding media" method by weakly absorbing backing. We show that for this kind of backing "variation of surrounding media" leads to completely wrong answer in determination of phase.

## II. DETERMINING THE PHASE BY VARIABLE BACKING METHOD

In the absence of significant non-specular scattering, neutron specular reflectometry is accurately described by a one-dimensional schrodinger equation for the neutron wave function. The exact wave function for an arbitrary finite film and its first derivative across the film can be carried from incident edge to transmitting edge by using the unimodular transfer matrix. If we consider non-vacuum fronting and backing, having constant SLD $\rho_f$ and $\rho_b$, respectively, using the continuity condition we have:

$$\begin{pmatrix} 1 \\ ib \end{pmatrix} t e^{ibqd} = \begin{pmatrix} A & B \\ C & D \end{pmatrix} \begin{pmatrix} 1+r \\ if(1-r) \end{pmatrix} \quad (1)$$

where q is normal component of the neutron wave vector in vacuum, d is the film thickness, (A,B,C,D) are the elements of transfer matrix that are function of the SLD of film and q. t and r are transmission and reflection coefficients respectively, and f and b are refractive indices of neutron for fronting and baking. Refractive index, n, for a sample having constant SLD, $\rho$, is related to q by

$$n^2 = 1 - 4\pi\rho/q^2 \quad (2)$$

The solution of Eq. (1) gives the reflection coefficient r, and reflectivity R,

$$r(q) = \frac{(f^2 b^2 B^2 + f^2 D^2) - (b^2 A^2 + C^2) - 2i(fb^2 AB + fCD)}{\Sigma + 2} \quad (3)$$

$$R(q) = |r(q)^2| = \frac{\Sigma - 2fb}{\Sigma + 2fb} \quad (4)$$

where

$$\Sigma = (f^2 b^2 B^2 + f^2 D^2) + (b^2 A^2 + C^2) \quad (5)$$

Having the real part of reflection coefficient Rer(q), and the reflectivity R, the imaginary part of reflection Imr(q), can be calculated from quadratic equation $|Rer(q)|^2 + |Imr(q)|^2 = R(q)$. This equation has two possible answers but only one satisfies Imr(q)→0 as q→0. For a free film, i.e., vacuum fronting and backing, Eqs. (3) and (4) become

$$r = \frac{(B^2 + D^2) - (A^2 + C^2) - 2i(AB + CD)}{\Sigma + 2} \quad (6)$$

$$r = \frac{(B^2 + D^2) + (A^2 + C^2) - 2}{(B^2 + D^2) + (A^2 + C^2) + 2} \quad (7)$$

Eqs. (6) and (7) show that $(B^2+D^2)$ and $(A^2+C^2)$ are sufficient to determine Rer(q) and R(q).

For a free film, with thickness d, the reflectivity does not change by using $\rho(z-d)$ instead of $\rho(z)$. It shows by extracting the SLD for mirror-reversed free film, the SLD of the free film can be determined. By exchanging, $\rho(z) \to \rho(z-d)$, A and D interchange in transfer matrix. So having {$B^2+D^2$ and $A^2+C^2$} or {$A^2+B^2$ and $C^2+D^2$}, we can extract the SLD of free film.

For the case of fronting medium is vacuum, Eq. (5) reduces to

$$\Sigma = b^2(A^2 + B^2) + (C^2 + D^2) \quad (8)$$

So by measuring two reflectivity spectra, i.e., two $\Sigma$ with different SLD, i.e., different b, we can calculate $(A^2+B^2)$ and $(C^2+D^2)$.

$$A^2 + B^2 = \frac{\Sigma_1 - \Sigma_2}{b_1^2 - b_2^2} \quad (9)$$

$$C^2 + D^2 = \frac{b_1^2 \Sigma_2 - b_2^2 \Sigma_1}{b_1^2 - b_2^2} \quad (10)$$

In Eqs. 1-10 the fronting and backing are considered as a semi-infinite backing ($R_\infty$), but what is measured in experiment is the reflectivity of thick backing ($R_{d\to\infty}$). The difference between $R_\infty$ and $R_{d\to\infty}$ may be appreciable for weakly absorbing backing. In next section we calculate $R_\infty$, $R_{d\to\infty}$, $\Sigma_\infty$ and $\Sigma_{d\to\infty}$ for two examples.

## III. DIFFERENCE BETWEEN THE REFLECTIVITY OF SEMI-INFINITE AND THICK BACKING

To show the difference between the reflectivity for thick and semi-infinite backing, we first treat the case of a thick film and then the general case of a thin film with constant SLD and thickness d and same fronting and backing, as an example of a thin film mounted on a thick backing. In the former, showing the refractive index for fronting and film by f and b, respectively, elements of transfer matrix are; $A=D=\cos(bq_0d)$ $C=-b^2B=\sin(bq_0d)/b$.

Using Eqs. (3) and (4), we find $\operatorname{Re}r(q)$, $\operatorname{Im}r(q)$ and $R(q)$ as given below

$$R(q) = \frac{f^2 - b^2}{f^2 + b^2} \operatorname{Re}r(q) = 1 - \frac{1}{1 + \xi^2 \sin^2(bqd)} \quad (11)$$

$$\operatorname{Im}r(q) = -\frac{\xi}{4} \frac{\sin(2bqd)}{1 + \xi^2 \sin^2(bqd)} \quad (12)$$

where

$$\xi = \frac{f^2 - b^2}{2fb} \quad (13)$$

In experiment we measure the average value of reflectivity. So here we must consider averages over the wave vector q. Assuming $\xi$ to be slowly varying over the averaging interval $\Delta \gg \pi/(bd)$, $\operatorname{Im}r(q)$ averages to zero and

$$\left\langle \frac{1}{1 + \xi^2 \sin^2(hqd)} \right\rangle = \sqrt{\frac{1}{1 + \xi^2}} \quad (14)$$

so

$$\langle r \rangle = r_{d \to \infty} = \frac{f - b}{f + b} \quad (15)$$

$$\langle R \rangle = R_{d \to \infty} = \frac{(f - b)^2}{f^2 + b^2} \quad (16)$$

Using reflection coefficient and reflectivity for semi-infinite backing, i.e.,

$$r_\infty = \frac{f - b}{f + b} \quad (17)$$

$$R_\infty = \left(\frac{f - b}{f + b}\right)^2 \quad (18)$$

Eqs. (15) and (18) show that the average of the complex reflection coefficient of a film with finite large thickness, $r_{d \to \infty}$, is indeed equal to the semi-infinite film $r_\infty$, but for the reflectivity, $R_{d \to \infty}$ and $R_\infty$ are different. The relative error is:

$$\frac{R_{d \to \infty} - R_\infty}{R_\infty} = \frac{2fb}{f^2 + b^2} \quad (19)$$

Eq. (19) shows that $R_{d \to \infty}$ is always greater than $R_\infty$ and the relative error increases with increasing the incident wave vector.

Fig. 1, shows the $R_{d \to \infty}$ and $R_\infty$ for an example in which fronting is vacuum and film has constant SLD, $\rho_b = 2.2 \times 10^{-4}$ nm$^{-2}$. Relative error for this example has been shown in Fig. 2. It is obvious that for large value of neutron incident wave vector $R_{d \to \infty}$ is twice as large as $R_\infty$.[14]

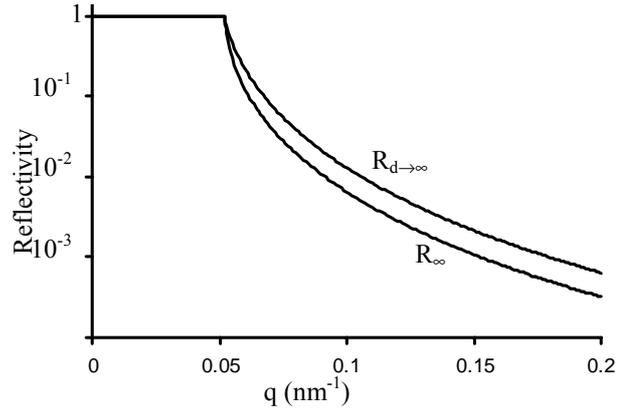

FIG. 1. $R_{d \to \infty}$ and $R_\infty$ for vacuum fronting and thick film having $\rho_b = 2.2 \times 10^{-4}$ nm$^{-2}$.

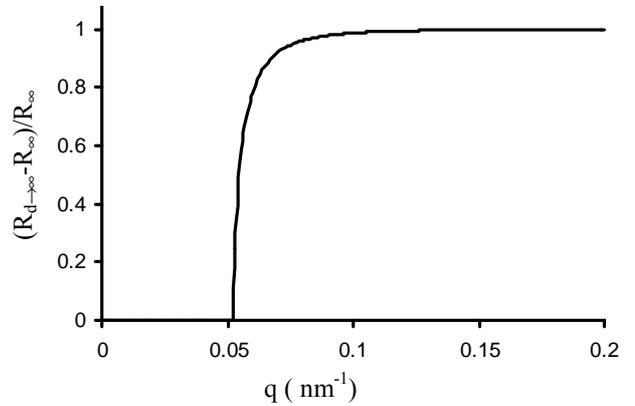

FIG. 2. Relative error for vacuum fronting and thick film having $\rho_b = 2.2 \times 10^{-4}$ nm$^{-2}$. It is obvious that for large value of incident neutron wave vector, $R_{d \to \infty} = 2R_\infty$

In the latter case, i.e., a thin film on a thick backing, the same calculation like for Eqs. (14) and (15), shows that the reflectivity for semi-infinite and thick backing are different too. In this case assuming as before that the reflection and transmission coefficients for free film, i.e., vacuum backing and fronting, are slowly-varying over the averaging interval, $R_{d\to\infty}$ is given by using the reflectivity for free:

$$R_{d\to\infty} = \frac{((\eta-1)\mu(1-R))^2}{\eta^2 - \mu^2(1-R)} \quad (20)$$

where

$$\eta = \frac{2n}{n-f} \quad (21)$$

and

$$\mu = \frac{f-b}{f+b} \quad (22)$$

where n is the refractive index for film and R is the reflectivity for free film. In Fig. (3-b) we show the difference between semi-infinite and thick backing for the arrangement shown in Fig. (3-a), a film with $\rho = 4\times 10^{-4}$ nm$^{-2}$, vacuum fronting and non-vacuum backing $\rho_b = 2.2\times 10^{-4}$ nm$^{-2}$. Fig. (3-c) shows the relative error for this example. The curve shows that the relative error has some extremum in the extremum of the reflectivity, i.e., the effect of thickness is more pronounced in the maxima of reflectivity.

## IV. EFFECT OF THICKNESS ON DETERMINATION OF PHASE

To determinate the phase of the reflection coefficient for a free thin film, we must use two $\Sigma(q)$ spectra corresponding to two measured reflectivity spectra with different values of the backing $\rho_b$ in Eqs. (9) and (10). These equations are obtained under the simplifying assumption that the backing is a semi-infinite matter. We show the effect of backing thickness for weakly absorbing backing on determination of the phase by replacing $\Sigma_{d\to\infty}$ for $\Sigma_\infty$. In Fig. (4) we plot $(\Sigma_{d\to\infty}-\Sigma_\infty)/\Sigma_\infty$, for the arrangement shown in Fig. (3-a). It is seen that the difference between these two cases is not so larg especially for large values of the incident neutron wave numbers. But if we use $\Sigma_{d\to\infty}$ in Eqs. (9) and (10) instead of $\Sigma_d$, the results for the real and imaginary parts of the reflection coefficient are completely wrong. Fig. (5) shows Rer(q) and Imr(q), derived from two reflectivity data with different values of the backing $\rho_b = 1\times 10^{-4}$ nm$^{-2}$ and $\rho_b = 2.2\times 10^{-4}$ nm$^{-2}$. In this figure the circles are recovered from using $\Sigma_\infty$ and the pluses from using $\Sigma_{d\to\infty}$. These two results show that for weakly absorbing materials as backing, variable surrounding method leads to completely wrong results.

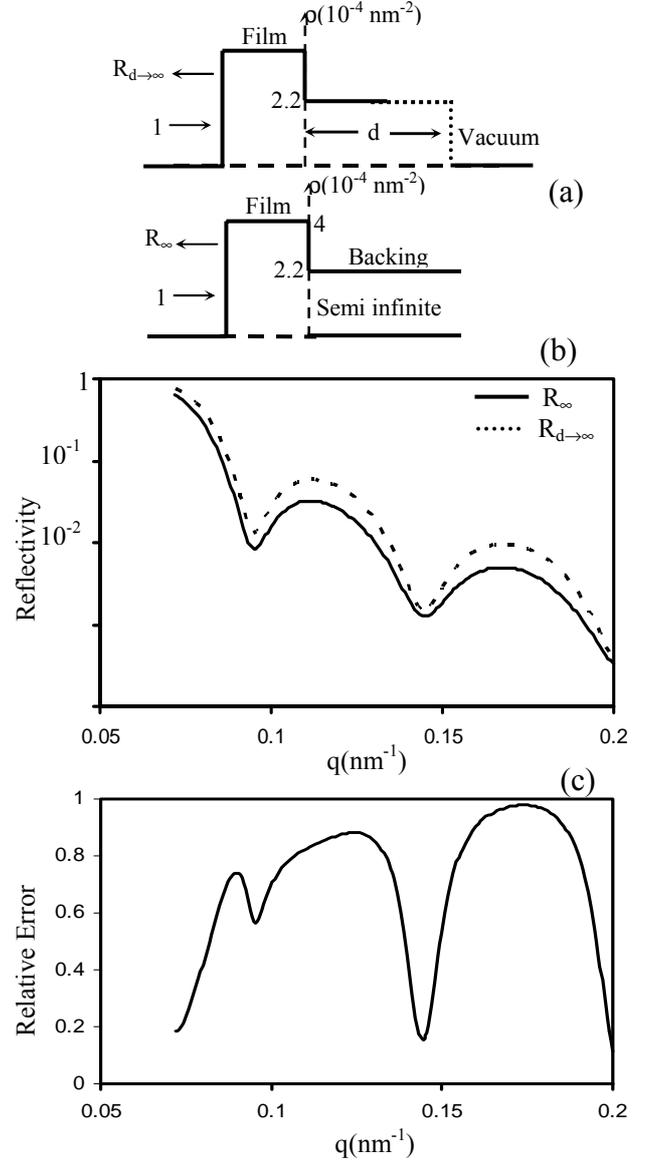

Fig. 3. (a) The arrangement used to show the difference between the reflectivity for semi-infinite and thick backing, shown in (b) by solid-line and dotted line respectively. (c) The relative error for this example. It shows that relative error has some extremum in the extremum of the reflectivities.

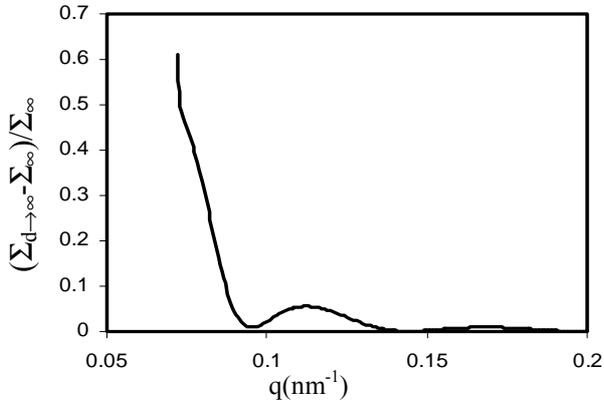

FIG. 4. $(\Sigma_{d\to\infty}-\Sigma_d)/\Sigma_\infty$ respect to the arrangement shown in Fig. (3-a).

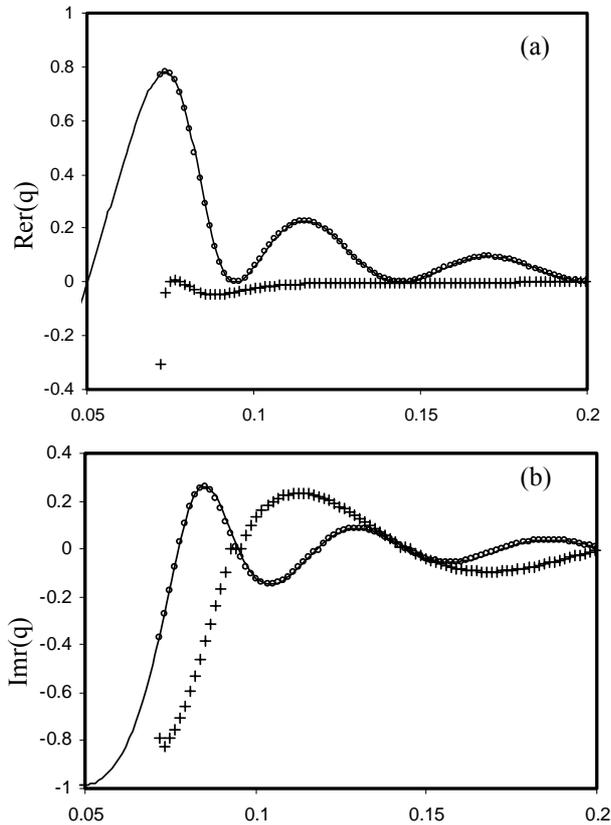

FIG. 5. (a) Rer(q). (b) Imr(q). Solid line: Computed directly from Eq. (6). Circles: recovered using reflectivity for semi-infinite backing. Plus: recovered using thick backing. Recovered data start at critical q of the thin film.

## VI. CONCLUSIONS

For weakly absorbing materials as backing like silicon, it is important to consider it as a thick matter instead of semi-infinite matter because of the reflection from the ends side of the backing causes very important effects on the reflectivity. We showed that taking into account these effects in "variation of surrounding media" method, for this kind of backing materials, leads to completely wrong answer in real and imaginary part of reflection coefficient that used to reconstruct the SLD profile which means that this method is not applicable for weakly absorbing backing.

## REFRENCES